\documentclass[journal]{IEEEtran}

\usepackage{graphicx}
\usepackage{xcolor}
\usepackage{listings}
\usepackage{svg}
\usepackage{comment}

\begin{document}
\title{Introducing Large Language Models as the Next Challenging Internet Traffic Source}
% critical, challenging, dominant, 


\author{Nataliia~Koneva,
        Alejandro~Leonardo~Garc\'{i}a~Navarro,
        Alfonso~S\'anchez-Maci\'an,~\IEEEmembership{Member,~IEEE,}
        Jos\'e~Alberto~Hern\'andez,~\IEEEmembership{Member,~IEEE,}
        %Jing~Fu,~\IEEEmembership{Member,~IEEE,}
        Moshe~Zukerman,~\IEEEmembership{Life Fellow,~IEEE,}
        \'{O}scar~Gonz\'alez~de~Dios% <-this % stops a space
        \thanks{Nataliia Koneva, Alejandro~Leonardo Garc\'{i}a Navarro, Alfonso S\'anchez-Maci\'an, and Jos\'e Alberto Hern\'andez are with the Department of Telematic Engineering, Universidad Carlos III de Madrid (UC3M), Spain.}
        \thanks{Moshe Zukerman is with the Department of Electrical Engineering, City University of Hong Kong.}
        \thanks{\'{O}scar Gonz\'{a}lez de Dios is with Telef\'onica I+D, Madrid, Spain.}}

\maketitle

\begin{abstract}
This article explores the growing impact of large language models (LLMs) and Generative AI (GenAI) tools on Internet traffic, focusing on their role as a new and significant source of network load. As these AI tools continue to gain importance in applications ranging from virtual assistants to content generation, the volume of traffic they generate is expected to increase massively. These models use the Internet as the global infrastructure for delivering multimedia messages (text, voice, images, video, etc.) to users, by interconnecting users and devices with AI agents typically deployed in the cloud. We believe this represents a new paradigm that will lead to a considerable increase in network traffic, and network operators must be prepared to address the resulting demands. To support this claim, we provide a proof-of-concept and source code for measuring traffic in remote user-agent interactions, estimating the traffic generated per prompt for some of the most popular open-source LLMs in 2025. The average size of each prompt query and response is 7,593 bytes, with a standard deviation of 369 bytes. These numbers are comparable with email and web browsing traffic. However, we envision AI as the next ``killer application" that will saturate networks with traffic, such as Peer-to-Peer traffic and Video-on-demand dominated in previous decades. 

% introduces the internet traffic of large language models (LLMs) as a new paradigm where we envision a massive deployment of AI Agents worldwide in the near future. These will interact with humans or other AI agents for healthcare, education, assistance, and many other activities that we cannot event think about at present. In general, all these agents will use the Internet as the global infrastructure for delivering multimedia messages (mainly text, but also voice, images, video, or programming code among others) to the users. We believe that this is a new paradigm that will fill networks with a considerable increase in traffic and network operators must be ready to address the new demands. We provide a proof-of-concept and source code to run traffic measurement experiments on remote User-Agent interactions and estimate the traffic generated per prompt for some of the most popular open-source Large Language Models in 2024. The average size of each prompt query and response is \textcolor{red}{xxxx} bytes with a standard deviation of \textcolor{red}{xxx} bytes. These numbers are comparable with email and web both in terms of queries and message sizes, so we envision that the Internet-of-Agents will contribute with similar percentages to the total traffic share, that is, with another 3-9\% over the total.
\end{abstract}


\IEEEpeerreviewmaketitle

\section{Introduction}

\IEEEPARstart{G}{enerative} Artificial Intelligence (GenAI) and Large Language Models (LLMs) appeared only a few years ago, yet they are already having a dramatic impact on society. These technologies demonstrate remarkable capabilities at working with text (including understanding and processing text, generating creative content such as songs and poems, or translation, among other functionalities), images (e.g. Dall-E~\cite{ramesh2021zeroshottexttoimagegeneration}, Midjourney~\cite{midjourney} and Stable Diffusion~\cite{rombach2022highresolutionimagesynthesislatent}), video (e.g. SORA~\cite{sora} and Flux~\cite{flux}), and music (e.g. MusicLM~\cite{agostinelli2023musiclmgeneratingmusictext}, Jukebox~\cite{dhariwal2020jukeboxgenerativemodelmusic}, Suno~\cite{suno} and Udio~\cite{udio}). 

The Internet of Things (IoT) paradigm emerged some years ago when the scientific community envisioned an Internet populated with connected ``things" beyond traditional computers or smartphones.  
These devices are primarily low-cost, battery-powered sensors and actuators with 5G/6G connectivity, massively deployed ubiquitously (in homes, rural areas, remote locations) to enable applications such as Smart Homes, Smart Agriculture, etc~\cite{CHOUDHARY2025100607}. 
This idea was later extended to ``Connected Industry 4.0", where manufacturers and industries integrate WiFi and 5G/6G connectivity into machines and robots~\cite{electronics13101832}. 
Similarly to IoT, the next step may involve an “Internet of Agents,” consisting of intelligent and collaborative agents augmenting IoT environments.  
In this vision, AI agents 
will massively populate the Internet, interacting not only with users and IoT devices, but also with other AI agents~\cite{LLM_Industry_commag}. 
These agents could operate locally on users' devices (such as laptops or smartphones) or be hosted remotely in the cloud, accessible via traditional web services like JSON over HTTP~\cite{LLM_telecoms}. 

For example, a specific Cardiology specialist AI agent may be running in the Cloud, and processing the user's biometric signals in real time. Placing this system in the cloud allows for capabilities such as real-time processing and enhanced privacy, as it is constantly retrained on new datasets collected worldwide from patients. In this particular case, there is continuous traffic exchange between the patient and the remote agent. This is just an example of a cloud-based medical AI application, but a wide range of similar services are expected to appear~\cite{Simbex2024, UPennAICloud2024}. 

Indeed, the deployment of intelligent agents is growing rapidly worldwide, though still in relatively early stages. A 2024 survey by Capgemini~\cite{capgemini2024generative} of executives at large enterprises reveals that 10\% of organizations are already using these systems, while over 50\% plan to explore using them within the next year. Furthermore, 82\% of surveyed organizations plan to integrate them within the next three years. This indicates that while widespread adoption is still limited, there is significant interest and planned growth in AI agent deployment over the next few years.

The market related to these AI agents is experiencing rapid expansion, with its size estimated at \$5.1 billion in 2024. 
Projections suggest substantial growth, with the market expected to reach \$47.1 billion by 2030~\cite{marketsandmarkets2024aiagents}. This represents an extraordinary compound annual growth rate (CAGR) of 44.8\% between 2024 and 2030, underscoring the surging importance and adoption across various industries and applications~\cite{Litslink2024}.

This article aims to quantitatively explore the impact on the telecommunications infrastructure of these systems. To this end, Section~\ref{sec:llms} introduces how Generative AI and LLMs have already impacted different fields. Section~\ref{sec:usecases} focuses on examples of services that AI agents may offer worldwide. Section~\ref{sec:poc} shows the methodology where we have evaluated how much traffic different query types to agents may account for. Section~\ref{Results} further explores the amount of traffic that examples of such queries may account for on the Internet. Finally, Section~\ref{sec:conclusion} concludes this work with a summary of its main findings and conclusions. 

%But also in scientific areas like Medicine, Biology, Physics, Computing and Coding, etc. 

%In this article, we introduce the Internet of Agents (IoA) as a new paradigm that is foreseen to appear soon (probably in months).

%We further envision a world massively populated with AI Agents connected to the Internet and communicating with Things, Users and other Agents. This includes examples of Cardiology-specialised agents answering questions to patients, or Travelling-agency agents monitoring trips and reporting them to users, Shopping-Agent assistants purchasing goods for homes or special educational agents as teachers for students.

%All these Agents communicating with other Agents or Humans or Things will contribute to increase network traffic at a pace above than current 40\% cumulative annual growth rate reported by Cisco Visual Networking Index and Ericsson Mobility Report.

%Many open-source GAI agents in LM studio: qwen (Alibaba), llama and llava (Meta), phi-3 (Microsoft), mixtral, etc

\section{Generative AI and Large Language Models}
\label{sec:llms}
% goal of this section: how Generative AI and Large Language Models have already impacted different fields

GenAI has revolutionized the field of artificial intelligence, marking a significant leap in machine learning capabilities. The foundation for this transformation was laid in 2017 with the publication of ``Attention Is All You Need"~\cite{vaswani2017attention}, which introduced the Transformer neural network architecture. This groundbreaking model, based entirely on attention mechanisms, eliminated the need for recurrence and convolutions, enabling more efficient parallel processing and superior performance in various language tasks. The Transformer architecture quickly became the backbone of LLMs, leading to the development of GPT (Generative Pre-trained Transformer) and its successors~\cite{OpenAI2023GPT4}. Further advancements came with the introduction of Reinforcement Learning from Human Feedback (RLHF), which allowed models to be fine-tuned based on human preferences, significantly improving their alignment with human values and intentions~\cite{yan2024rewardrobust}.

The release of ChatGPT 3.5 by OpenAI in late 2022 marked a turning point in the accessibility and capabilities of generative AI, sparking intense competition among tech giants and startups alike. Google responded with Gemini~\cite{geminiteam2024geminifamilyhighlycapable}, a multimodal AI model designed to understand and generate text, images, and code. Anthropic's Claude~\cite{anthropic2025claude} emerged as a strong contender, focusing on safety and ethical considerations. Mistral AI~\cite{mistral2025models}, a French startup, gained attention with its efficient and powerful open-source models. Perplexity AI~\cite{perplexity2025} aimed to revolutionize search with its AI-powered engine, while Elon Musk's xAI introduced Grok~\cite{grok2023}, emphasizing real-time knowledge and wit. Meta's LLaMA~\cite{touvron2023llamaopenefficientfoundation}, Alibaba's Qwen~\cite{bai2023qwentechnicalreport} and Deepseek~\cite{deepseekai2025deepseekr1incentivizingreasoningcapability} joined the race, offering open-source alternatives and specialized models for various applications. This competitive landscape has driven rapid innovation, with each player striving to enhance model performance, efficiency, and unique features. A good comparison of LLMs performance at different tasks involving text can be found in~\cite{zhang2024benchmarking}.

Emerging applications of generative AI have expanded beyond text, venturing into multimedia domains. OpenAI's Sora represents a significant breakthrough in video generation, capable of creating realistic and imaginative video content from text descriptions. Runway's Gen-2~\cite{runway2023gen2} and Stability AI's Stable Video Diffusion~\cite{blattmann2023stablevideodiffusionscaling} also push the boundaries of AI-generated video. Concerning image generation, Midjourney has gained popularity for its ability to create highly detailed and artistic images from text prompts, competing with models like DALL-E and Stable Diffusion. Music generation has seen advancements with models like Google's MusicLM and OpenAI's Jukebox, capable of creating original compositions across various genres. Code generation tools, such as GitHub Copilot~\cite{github2023copilot} and OpenAI's Codex~\cite{chen2021evaluatinglargelanguagemodels}, transform software development by assisting programmers with code completion and generation. These diverse applications demonstrate the versatility and potential of generative AI to revolutionize creative processes across multiple industries.

In this sense, the popularity of Generative AI applications has seen exponential growth in terms of user engagement and query volume. ChatGPT, the leading generative AI platform, reportedly receives over 1.6 billion visits per month, with users submitting an estimated 100 million queries daily. Other popular platforms, like Google's Gemini and Anthropic's Claude are also experiencing rapid user adoption, though specific figures are not publicly available. 

The demand for AI-powered services has led to a massive expansion in computing infrastructure. It is estimated that training LLMs requires hundreds of thousands of GPUs. For instance, OpenAI's GPT-3 was trained on a supercomputer cluster with over 285,000 CPU cores and 10,000 GPUs. Major tech companies and AI research organizations are rapidly expanding their data center capabilities, with some reports suggesting that the total number of GPUs dedicated to AI training and inference could exceed 10 million by 2025. This infrastructure growth underscores the increasing computational demands of generative AI and the industry's commitment to meeting the surging user interest in these technologies.

%Citations:
%[1] https://www.flexos.work/learn/generative-ai-top-150
%[2] https://www.salesforce.com/news/stories/generative-ai-statistics/?bc=DB
%[3] https://bloggingwizard.com/generative-ai-statistics/
%[4] https://www.ciodive.com/news/top-generative-ai-apps-for-employees/704371/
%[5] https://www.mckinsey.com/capabilities/quantumblack/our-insights/the-state-of-ai-in-2023-generative-ais-breakout-year
%[6] https://futureskillsacademy.com/blog/top-generative-ai-applications/
%[7] https://www.coursera.org/articles/generative-ai-applications
%[8] https://www.turing.com/resources/generative-ai-applications

\section{Internet of Agents: Use cases and applications}
\label{sec:usecases}

\begin{figure}
    \centering
    \includegraphics[width=0.9\linewidth]{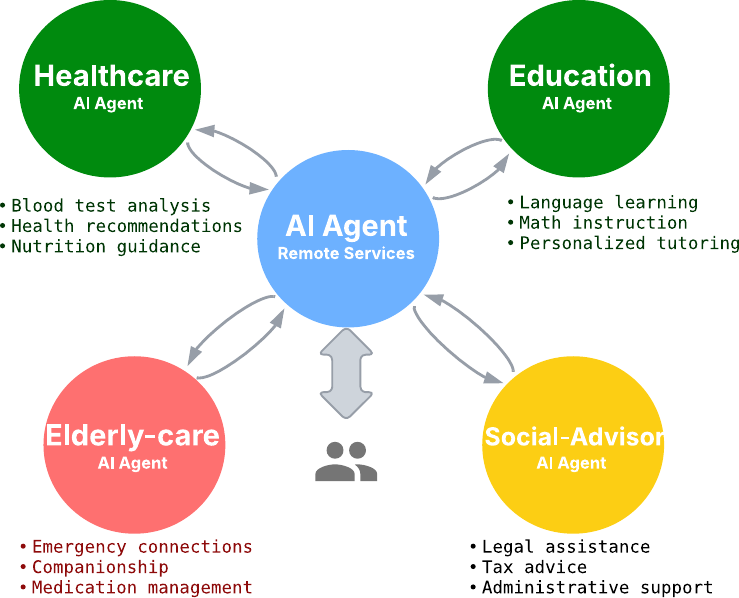}
    \caption{Use-cases and applications of AI Agents in future life}
    \label{fig:usecases}
\end{figure}

In a nutshell, we envision an Internet of Agents (IoA) developed by integrating AI technologies with specialized datasets, where specific AI agents running on the cloud interact with users, sensors, connected devices, and other AI agents. 
The following are examples of use cases and applications of such an IoA (see also Fig.~\ref{fig:usecases}).

\textit{Healthcare AI Agents} have been massively trained with healthcare data sets and have specific instructions to answer healthcare-related queries. For example, we can have an agent that can receive a document with a blood test from a user and explain it to the user, giving further recommendations regarding healthy habits to improve his/her health conditions. The same applies to Nutrition-based agents, who can assist customers with healthy recipes for the next meal. 

\textit{Elderly-care AI Agents} can provide valuable social assistance to the elderly in multiple ways, including emergency connections, interactions with police and doctors, and shopping support. As an example, AI assistants can be activated via voice commands, allowing them to recognize emotions in voice, assess the situation, and contact emergency services or family members in case of an emergency. Beyond emergency support, AI agents can help combat loneliness by engaging in conversations and games, managing medication schedules with reminders, and monitoring health via integration with wearable devices. %By combining these functionalities, the AI agent serves as a comprehensive support system, improving the quality of life for elderly individuals while providing peace of mind to their families and caregivers.

\textit{Education AI Agents} are transforming education by providing personalized support in language learning, tutoring, and math instruction. They can tailor lessons to individual students' proficiency levels, offer interactive conversation practice, and deliver instant feedback on exercises. AI can generate customized practice problems in mathematics, provide step-by-step solutions, and adapt performance-based difficulty. This personalized approach enhances engagement and helps students grasp complex concepts more effectively. Additionally, AI agents can take on roles as philosophers or religious figures, facilitating a deeper understanding of humanities subjects. By simulating dialogues with historical thinkers, they make complex ideas more accessible and encourage critical thinking. The benefits of using AI in education include 24/7 availability, consistent quality of instruction, and data-driven insights into student performance. %However, it is essential to implement these tools ethically, ensuring they complement human teachers and protect student privacy while delivering high-quality educational experiences.

\textit{Social-advisor AI Agents} provide specialized assistance in fields such as law, taxation, and administration. In the legal domain, AI lawyer advisors can assist in drafting contracts by analyzing vast databases of legal documents and precedents, ensuring that all necessary clauses are included and tailored to specific requirements. They can also interpret complex laws and regulations, providing advice and comprehensive summaries to the non-expert. This also applies to taxation and administration, where AI agents can provide advice, help with legal document filing, and streamline various administrative processes, such as reading invoices and providing structured summaries of accounting documents.

%[nice powerpoint figure ]
%Here a nice powerpoint figure with some uses and applications

For this, humans who interact with AI agents will increase the amount of traffic that traverses the network as it is inexpensive for users. Devices like Alexa, Cortana, and Siri from Amazon, Microsoft, and Apple, respectively, will be so accessible that it will be very straightforward to interact with them. 

The next sections aim to characterize and estimate the amount of network traffic increased due to the Internet generated by such agents.

\section{Experiments}
\label{sec:poc}

\subsection{Proof of Concept and Scenario}
\label{concept}

To better understand the impact of AI Agents on future Internet traffic, we designed a proof-of-concept experiment to quantify the amount of network traffic generated during typical interactions between users (or other agents) and LLMs. While previous sections explored potential applications and user growth trends, this section focuses on a practical, measurement-based approach to assess real-world traffic implications. Specifically, we aim to characterize how much data is exchanged per prompt when querying widely-used LLMs.

The setup simulates an interaction between two AI agents during prompt-response interactions,  where two AI agents (querying agent and responding agent) communicate over a local machine. The querying agent fetches input prompts from a benchmark set and sends them as HTTP requests to the Responding Agent. The Responding Agent is a local server that forwards the queries to various LLM APIs, such as LLaMA and OpenAI, and then returns responses.

The experiment evaluates seven LLMs: MistralAI (mistral-7b), Gemini-pro, Llama3.1-70b, Llama3.2-11b-vision, qwen-2.5-32b via Groq, OpenAI (gpt-3.5-turbo-instruct), and DeepSeek (deepseek-r1:7b). In Table~\ref{tab:llm_overview}, we summarize the default inference settings and model specifications used in our experiments. The \textit{Param.} column refers to the approximate number of trainable parameters in the model, expressed in billions (B). The \textit{Context Length} defines the maximum number of tokens a model can process in a single forward pass, encompassing both input (prompt) and output (response). The \textit{Max Tokens} field specifies the default maximum number of tokens generated in a single response, which can often be customized but defaults to fixed values (e.g., 4,096 or 1,024) in most API implementations. \textit{Temperature} controls the randomness of the generated text: lower values yield more deterministic outputs, while higher values promote diversity and creativity. Finally, \textit{Top-p} refers to nucleus sampling, which selects tokens only from the smallest subset whose cumulative probability exceeds a threshold (e.g., 0.9 or 1.0), balancing coherence and variation. %These default values were either explicitly set or inferred from official documentation of each model’s provider.

% JOSE TABLE Name Company Date of Release Parameters

\begin{table*}[!htbp]
\centering
\caption{Overview of Evaluated LLMs and Default Inference Settings}
\begin{tabular}{|l|l|l|l|l|l|l|l|}
\hline
\textbf{Model Name} & \textbf{Company} & \textbf{Date of Release} & \textbf{Param}. & \textbf{Context Length} & \textbf{Max Tokens} & \textbf{Temp.} & \textbf{Top-p}\\
\hline
MistralAI (open-mistral-7b) & Mistral & Sept. 2023 & 7B & 8,192 & 4,096 & 0.7 & 1.0 \\ %vocab_size	32000

Claude-3-sonnet & Anthropic & Mar. 2024 & N/A & 200,000 & 4,096 & 0.7 & 1.0 \\

LLaMA3.1–70B & Meta & Apr. 2024 & 70B & 128,000 & 4,096  & 0.7 & 0.9 \\

LLaMA3.2–11B-Vision & Meta & Sept. 2024 & 11B & 128,000 & 4,096 & 0.7 & 0.9 \\

Qwen-2.5-32b (Groq) & Alibaba & Sept. 2024 & 32B & 128,000 & 1,024 & 0.7 & 1.0 \\

GPT-3.5-turbo-instruct & OpenAI & Mar. 2023 & 175B & 16,385 & 4,096 & 0.9 & 1.0 \\

DeepSeek R1 & High-Flyer AI & Jan. 2025 & 7B & 128,000 & 32,768 & 0.8 & 0.9 \\

\hline
\end{tabular}
\label{tab:llm_overview}
\end{table*}

Using Wireshark, we capture and analyze two types of traffic: the communication between the querying agent and responding agent, and the traffic between the Responding Agent and the external LLM APIs.

\subsection{Traffic Measurement Methodology and Source-Code}
\label{Methodology}

To analyze the traffic patterns in more detail, we utilized the OpenAI dataset GSM8K from \cite{cobbe2021trainingverifierssolvemath}, which provides a diverse collection of approximately 8,500 linguistically diverse grade school math problems. This dataset was particularly useful for our analysis, as it contains structured JSON files with varying complexity levels of interactions. For this experiment, we selected 1,000 questions from a benchmark dataset and adapted the mathematical problems into agent-human dialogues to simulate realistic conversational patterns while maintaining the structured nature of the interactions. Each question was sent as an individual request, and the corresponding responses were systematically collected. The traffic capture process involved assigning a unique stream index in Wireshark to each query-response pair. 

%Consistent with initial observations, the responses were invariably segmented into three TCP segments for each LLM model. The first and third segments maintained consistent sizes of 184 bytes and 49 bytes, respectively, while the second segment exhibited notable variations in length across different models. 

The code for this experiment has been uploaded to GitHub~\cite{LLMTrafficMonitor} and can be downloaded for re-running these experiments and upgrading to new scenarios. The code is written in the open-source programming language \textit{Python} and utilizes several key libraries to enable agent communication. Specifically, the \textit{Flask} framework establishes the responding agent as an HTTP server, allowing it to receive queries from the querying agent and forward them to various LLM APIs. The responding agent incorporates the \textit{langchain-openai}, \textit{ollama}, or \textit{openai} library, which simplifies API calls to retrieve model-generated responses.

%screenshot of the wireshark (localhost and wifi) + d filtering + stream index

To analyze the network traffic generated by LLM-based interactions, we captured and examined both localhost traffic (exchange locally between agents) and external network traffic (traffic between the local server and remote, cloud-hosted LLM APIs) using Wireshark. Each question was assigned a unique Stream Index in Wireshark, enabling accurate summation of all packet lengths associated with a single query-response pair. This allows us to compute the total request and response size per LLM in both localhost and external network traffic scenarios.

\begin{figure}[!htbp]
    \centering
    \includegraphics[width=\linewidth]{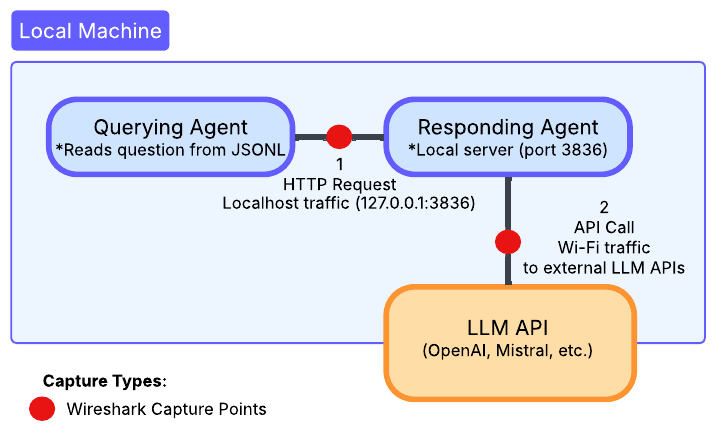}
    \caption{Experimental setup and packet capture points}
    \label{fig:experiment-diagram}
\end{figure}

Wireshark captures data at two points, as illustrated in Fig.~\ref{fig:experiment-diagram} of the experiment setup. First, it records localhost traffic exchanged between the querying and responding agents. Second, it captures Wi-Fi traffic consisting of requests and responses between the responding agent and the external LLM APIs. 

\begin{figure*}[htbp]
    \centering
    \begin{minipage}[t]{1\textwidth}
        \centering
        \includegraphics[width=0.8\linewidth]{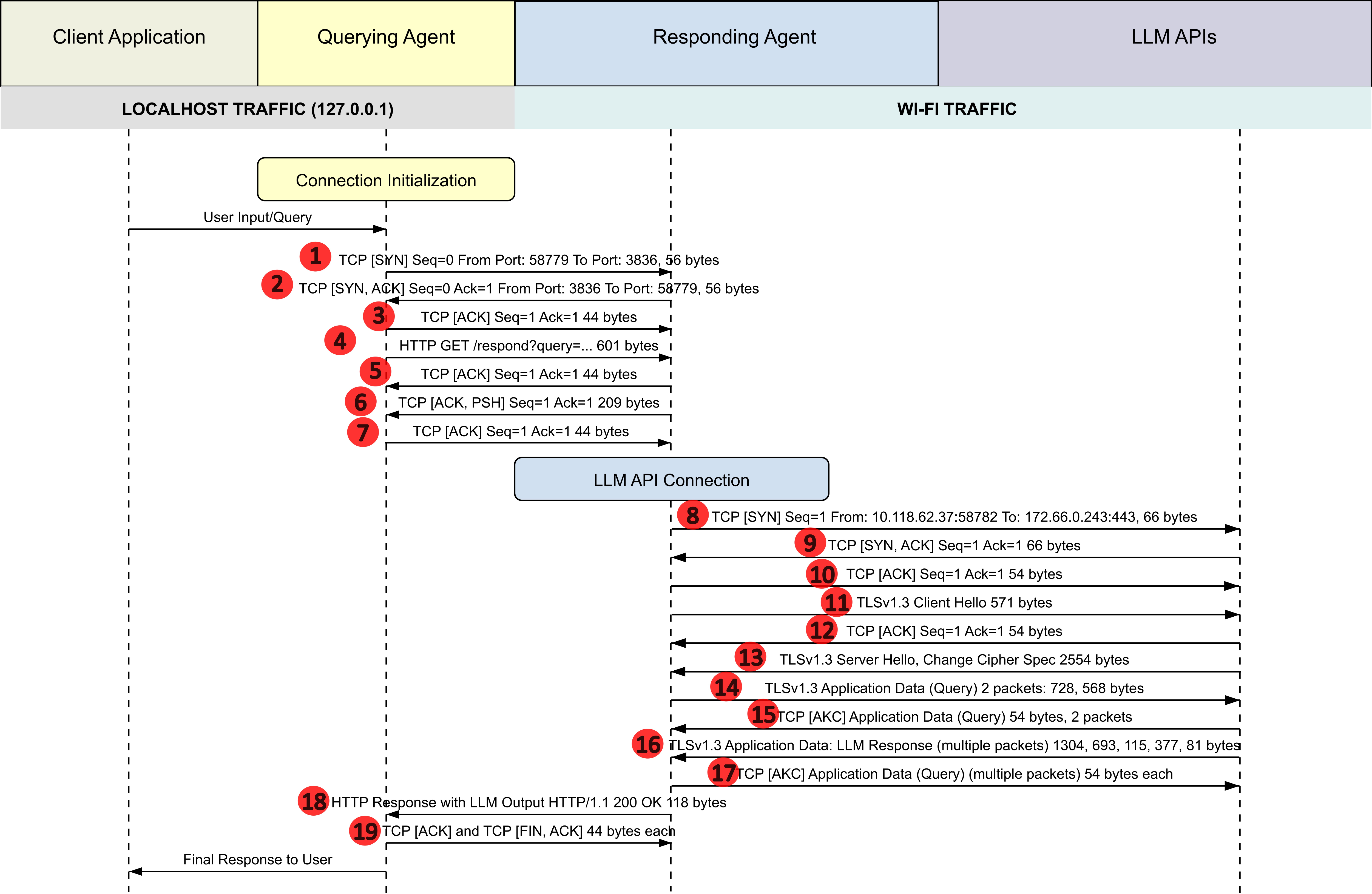}
        \caption{(a) LLM traffic workflow showing localhost and Wi-Fi packet exchanges}
        \label{fig:workflow}
    \end{minipage}\hfill
    \begin{minipage}[t]{\textwidth}
        \centering
        \includegraphics[width=0.8\linewidth]{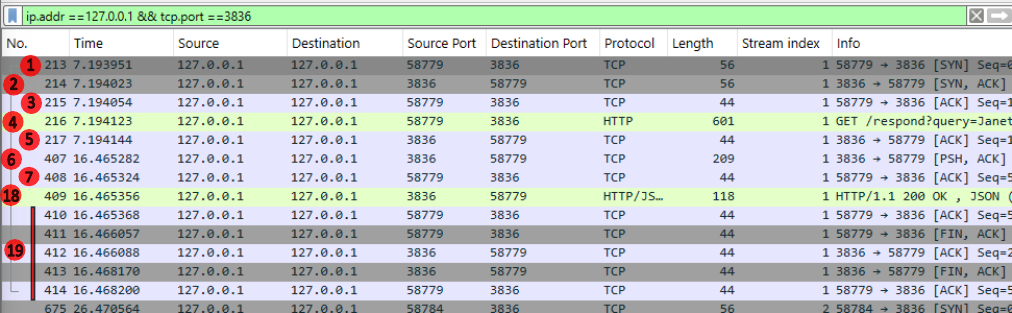}
        \caption{(b) Localhost traffic capture from Wireshark between Querying and responding agents}
        \label{fig:wireshark_localhost}
    \end{minipage}\hfill
    \begin{minipage}[t]{\textwidth}
        \centering
        \includegraphics[width=0.8\linewidth]{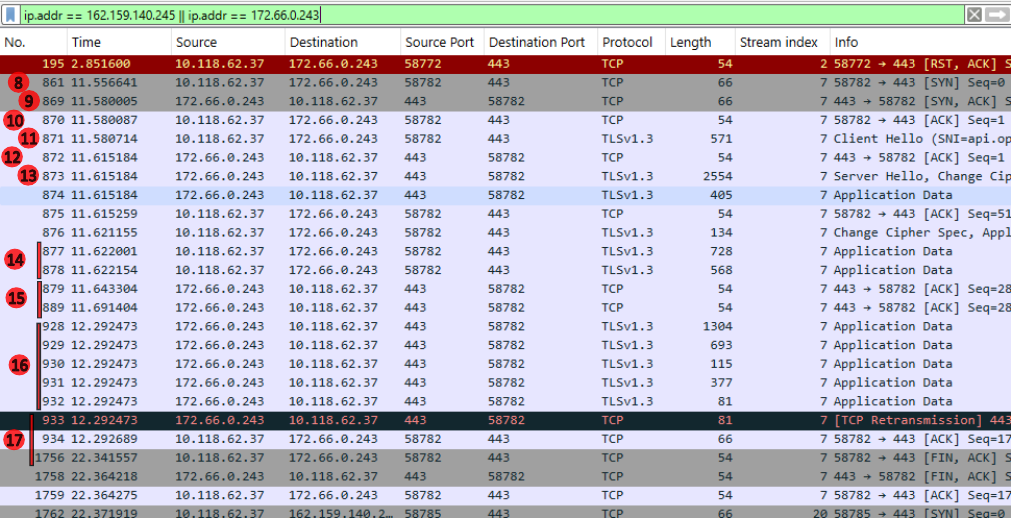}
        \caption{(c) External network traffic capture from Wireshark between Responding Agent and LLM APIs}
        \label{fig:wireshark_wifi}
    \end{minipage}
    \label{fig:combined_traffic}
\end{figure*}

The screenshot in Fig.~\ref{fig:wireshark_localhost} shows a captured localhost (127.0.0.1:3836) request-response exchange, where the total bytes exchanged per query were measured. The external traffic is captured using applied display filters to isolate the traffic of LLM providers from other traffic. Fig.~\ref{fig:wireshark_wifi} presents the captured Wi-Fi traffic for external API communication, showing individual requests sent to LLM APIs and their corresponding response sizes.

To provide a complete visualization of the traffic flow in our experiment, Fig.~\ref{fig:workflow} illustrates the end-to-end packet exchange between system components. On the left-hand side, it shows the localhost traffic between the querying agent and the responding agent, while the right-hand side represents the Wi-Fi traffic between the responding agent and the external LLM API servers. 
 
The communication begins with the user sending a query to the local server, triggering a standard TCP handshake (steps 1–3), followed by an HTTP GET request (step 4) from the querying agent. This message is acknowledged and passed to the Responding Agent (steps 5–7), which then initiates a connection to the external LLM API (steps 8–9).

A full TLS 1.3 handshake ensues (steps 10–13), after which the LLM API receives the prompt as encrypted application data (step 14) and returns the response (step 15), also split into multiple packets. This response is acknowledged (steps 16–17) and returned to the local querying agent (step 18), which then delivers the final output to the user (step 19).

This diagram helps identify where traffic is generated in each part of the system and serves as a reference when analyzing packet captures in Wireshark (as detailed in Figs.~\ref{fig:wireshark_localhost} and~\ref{fig:wireshark_wifi}).

 By summing the lengths of all packets in a given stream, we obtained the total bytes sent and received per query for each LLM, separately for the traffic of the internal agent communication and the external cloud-hosted API traffic. In essence, for each of the 1,000 queries (per model), we recorded two numbers: the bytes exchanged between agents on localhost, and the bytes exchanged with the LLM’s server over the network. These measurements provide the raw data for our traffic analysis.

\subsection{Results}
\label{Results}

\begin{figure*}[htbp]
    \centering
    \begin{minipage}[t]{1\textwidth}
        \centering
        \includegraphics[width=1\linewidth]{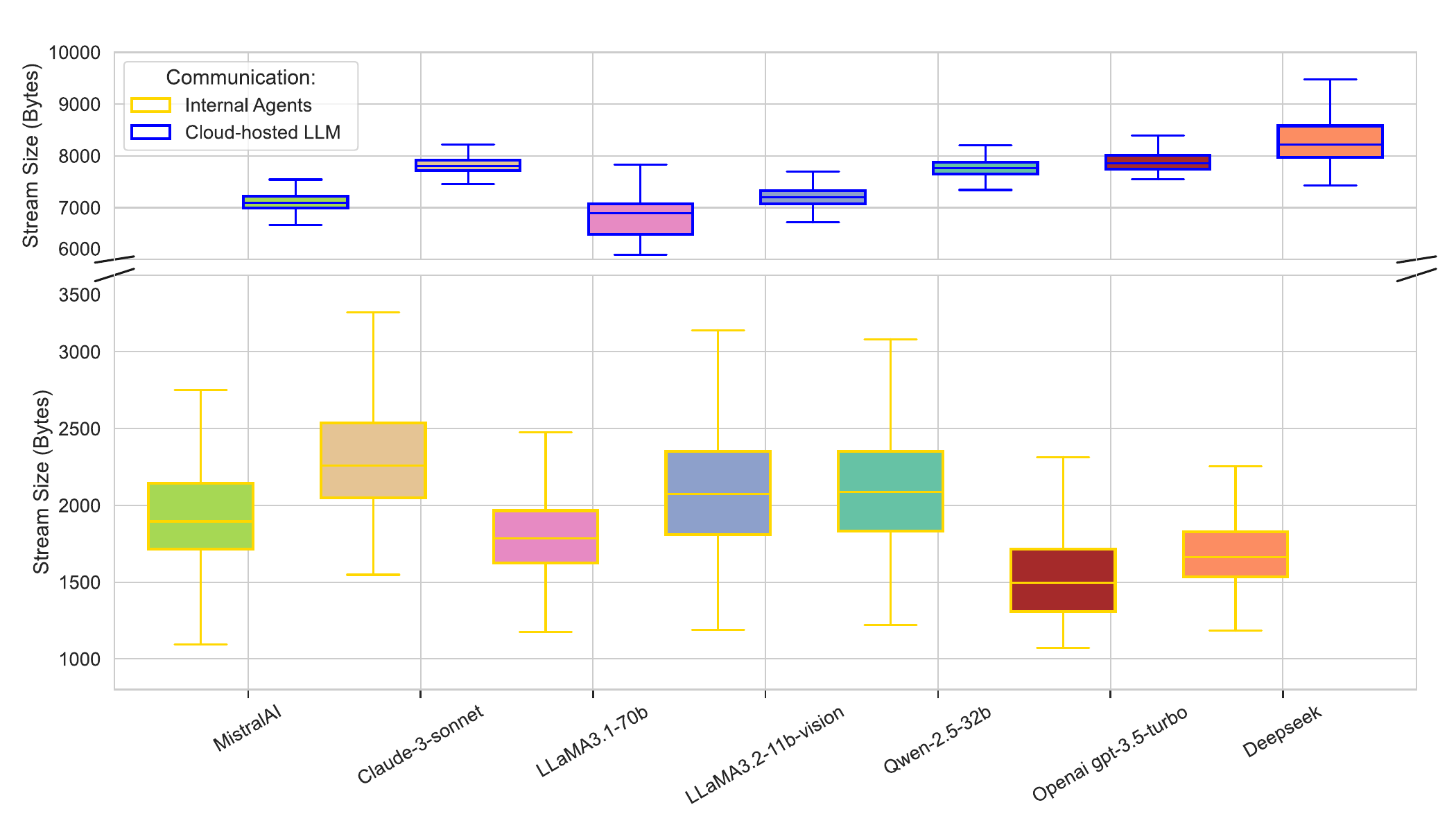}
        \caption{Distribution of Localhost and WiFi Traffic per LLM}
        \label{fig:workflow}
    \end{minipage}\hfill
    \label{fig:boxplot_wifi_localhost}
\end{figure*}

After executing the prompt-response trials for all seven LLMs and capturing the data, we analyzed the distribution of traffic per interaction and computed summary statistics. Fig.~\ref{fig:boxplot_wifi_localhost}  visualizes the distribution of total bytes exchanged per query (request + response) for each model, distinguishing between internal agent communication (local interaction between agents) and external API communication (requests sent over the Internet to cloud-hosted LLMs). DeepSeek exhibits minimal local traffic yet generates substantial external traffic, pointing to protocol and session management overheads. This discrepancy is likely due to frequent TLS handshakes, which introduce significant overhead in encrypted connections.  OpenAI, on the other hand, maintains efficiency across both communications, reflecting its efficient session reuse and minimal metadata exchange. %In Fig.~\ref{fig:boxplot_localhost}, which uses a logarithmic scale for clarity, we observe the spread of local traffic sizes for each LLM. 
%It highlights differences in both typical values and variability: for example, some models have a tight clustering of points (indicating consistent response sizes), while others show long tails or wider spread (indicating occasional much larger responses). %Such visualizations provided an initial qualitative confirmation of our expectations; for instance, larger models tended to exchange more data on average, and the variation in response size differed from model to model.
\begin{comment}
    \begin{figure}[!htbp]
    \centering
        \begin{minipage}[b]{\columnwidth}
        \centering
        \includegraphics[width=\textwidth]{fig/boxplot_localhost_v2.png}
        \caption{Distribution of Localhost Traffic per LLM in Logarithmic Scale}
        \label{fig:boxplot_localhost}
    \end{minipage}\hfill
    \begin{minipage}[b]{\columnwidth}
        \centering
        \includegraphics[width=\textwidth]{fig/boxplot_wifi_v2.png}
        \caption{Distribution of WiFi Traffic per LLM}
        \label{fig:boxplot_wifi}
    \end{minipage}\hfill
\end{figure}
\end{comment}

%results analysis for the default settings of the llm
%Average Localhost Request-Response Size: XXX bytes (std: XXX bytes)
%Average Wi-Fi Request-Response Size: XXX bytes (std: XXX bytes)
%Most efficient model in terms of traffic: (e.g., OpenAI GPT-3.5-turbo-instruct)
%Highest traffic model: (e.g., DeepSeek-r1:7b due to large response size)

The collected traffic data from the querying and responding agents was analyzed to determine the network footprint of different LLMs. The results are summarized in Tables~\ref{tab:benchmark_Localhost} and~\ref{tab:benchmark_Wifi} with mean, standard deviation, median, and quartile statistic parameters, which present the measured traffic for localhost and Wi-Fi transmissions, respectively. All models were run with default settings, ensuring a fair comparison. 

%Since DeepSeek (deepseek-r1:7b) runs only locally on Ollama, in the following sections, the data presented in Tab.~\ref{tab:benchmark_Localhost} and Tab.~\ref{tab:benchmark_Wifi} show that DeepSeek's network traffic was exclusively captured under localhost conditions, with no corresponding external network traffic measurements.

%Table~\ref{tab:benchmark} shows the results after measuring the query and responses for more than 3 thousand different prompts submitted to 8 LLM models following the methodology above. For the prompts, we used Microsoft's open-source prompt benchmark~\cite{zhu2024promptbenchunifiedlibraryevaluation}.

\begin{table*}[!htbp]
\centering
\caption{Traffic measured in bytes in (Local traffic between agents)}
\label{tab:benchmark_Localhost}
\begin{tabular}{|l|lllllll|}
\hline
\textbf{} & \textbf{Min} & \textbf{1st-Q} & \textbf{Median} & \textbf{Avg} & \textbf{3rd-Q} & \textbf{Max} & \textbf{Sd} \\ \hline
MistralAI & 1094.00 & 1714.75 & 1895.50 & 1961.18 & 2143.00 & 4873.00 & 398.37 \\
Claude-3-sonnet-20240229 & 1548.00 & 2047.50 & 2257.50 & 2305.60 & 2537.00 & 3724.00 & 359.68 \\
llama3.1-70b & 1144.00 & 1501.50 & 1792.00 & 1837.18 & 2100.75 & 7518.00 & 464.88 \\
llama3.2-11b-vision & 1191.00 & 1810.00 & 2074.00 & 2131.41 & 2350.25 & 7558.00 & 502.24 \\
Qwen-2.5-32b (Groq) & 1222.00 & 1833.00 & 2086.50 & 2120.99 & 2350.25 & 4266.00 & 389.84 \\
Openai gpt-4o & 1071.00 & 1310.00 & 1496.50 & 1546.78 & 1716.25 & 2748.00 & 297.75 \\
DeepSeek R1 & 1184.00 & 1536.00 & 1664.00 & 1702.52 & 1828.25 & 2660.00 & 235.23 \\ \hline
\end{tabular}
\end{table*}

\begin{table*}[!htbp]
\centering
\caption{Traffic measured in bytes (External API traffic)}
\label{tab:benchmark_Wifi}
\begin{tabular}{|l|lllllll|}
\hline
\textbf{} & \textbf{Min} & \textbf{1st-Q} & \textbf{Median} & \textbf{Avg} & \textbf{3rd-Q} & \textbf{Max} & \textbf{Sd} \\ \hline
MistralAI & 6668.00 & 6994.00 & 7092.00 & 7159.72 & 7219.00 & 9066.00 & 313.28 \\
Claude-3-sonnet & 7461.00 & 7713.00 & 7803.00 & 7834.97 & 7916.25 & 10074.00 & 216.59 \\
llama3.1-70b & 6095.00 & 6489.75 & 6814.16 & 6814.16 & 7077.25 & 7832.00 & 338.55 \\
llama3.2-11b-vision & 6673.00 & 7074.75 & 7202.00 & 7239.43 & 7323.00 & 14412.00 & 486.32 \\
Qwen-2.5-32b (Groq) & 7290.00 & 7648.75 & 7764.00 & 7794.98 & 7878.25 & 9480.00 & 250.94 \\
Openai gpt-4o & 7547.00 & 7740.00 & 7860.00 & 7885.38 & 8008.00 & 8500.00 & 179.57 \\
DeepSeek R1 & 7433.00 & 7971.75 & 8214.00 & 8419.37 & 8580.25 & 15901.00 & 799.37 \\ \hline
\end{tabular}
\end{table*}
A few clear trends emerge from these results. Models such as DeepSeek (deepseek-r1:7b) and qwen-2.5-32b via Groq exhibited the highest variability in network usage (reflected by their larger standard deviations).

OpenAI’s GPT-3.5-turbo-instruct consistently generated smaller response sizes compared to models such as Llama3.1-70b and Gemini-pro. GPT-3.5-turbo model is designed for efficient tokenization and optimized response length. This suggests that the responses from these models can vary greatly in length or content, leading to inconsistent traffic per prompt. 

On the other hand, OpenAI’s GPT-3.5-turbo-instruct consistently produced smaller response sizes on average compared to models like LLaMA-70B or Gemini-pro. This aligns with the model’s design goal of optimizing token efficiency and limiting response length for practical use. In contrast, the larger LLaMA-based models and Gemini-pro often returned more extensive answers, contributing to higher average bytes per interaction. Across all the LLMs tested, the average total traffic per prompt (request + response) was about 7.59 kB, with a standard deviation of roughly 0.37 kB. 

In other words, a typical question-answer exchange with a modern LLM involves on the order of only a few kilobytes of data transfer. This magnitude is comparable to the size of an average email or a small webpage, reinforcing our perspective that LLM-driven agent communication could become a notable component of Internet traffic. In fact, if AI agent interactions (the IoA paradigm) grow in popularity, they could significantly contribute to the overall Internet traffic volumes.

% av 1119.95 B in the localhost,  sd 1126.77 B
% av 3979.97 B in the wifi , Sd 2028.38 B

\section{IoA traffic forecasts}

\begin{comment}
As shown, the models that generate more traffic per query are xxxxxxx

%Ptotocols involved: Web services etc

%QUery 1, query 2, query 3, table with results in bytes

\section{On traffic growth estimates}
\label{sec:estimates}

The Internet is estimated to carry 33 Exabytes of data per day. While exact projections vary, experts generally agree that Internet traffic will continue to grow at around 20-25\% per year.

The majority of this traffic belongs to Video-on-Demand (65\% percentage), while other uses and applications comprise less traffic share, namely email accounts for 3\%, web accounts for 15\% and the Internet-of-Things 7\%, etc. In particular, email and web browsing are typically text-based where users and servers exchange text data in some format (messages, HTML, JS code, etc). In the case of IoT and M2M scenarios, very often the exchange is in the form of JSON messages. 

Taking as reference email traffic, there are 4.3 billion users, while the average number of emails per person is about 120-130 per day. The contents of each email are 75-100 words on average. Regarding web traffic, there are 1.9 billion websites on the Internet with average size of 2-3 MB.

We can take these examples (email, web) as starting estimates for Agent-based traffic since we expect AI Agents to exchange information in the form of text with users (with structured JSON format for instance).

To this end, Table~\ref{tab:ai_traffic_estimates} shows multiple scenarios regarding AI Agents popularity use and related traffic estimates. 
\end{comment}

At present (April 2025), the most visited websites globally, ranked by monthly visits, are Google, Youtube and Facebook, accounting for around 140, 78 and 12 billion visits per month respectively. According to similarweb~\cite{similarweb2024openai}, OpenAI and ChatGPT received 551.1 million total visits in September 2024, which is a 22.96\%  increase compared to the previous month. OpenAI ranks \#90 webpage with most traffic worldwide.

%As of March 2024, Perplexity AI received 52.4 million visits per month. This represents a 23.79\% increase from February 2024, when it received 42.33 million visits. 

The growth rate of visits per month shows an increase of about 20-24\% per month, which is approximately 9-fold per year, for the most popular AI websites (OpenAI, Anthropic, Mistral, etc). Based on the most recent data available as of October 2024, ChatGPT currently has over 180 million registered users worldwide, and most of them are active every week.

To estimate the long-term impact of AI agents on network traffic, we can create a simple traffic forecast model based on three parameters: the number of users, the average size per response, and the average queries per user per day. We assume that each interaction or visit to an AI agent represents one request-response exchange.

According to recent reports, OpenAI’s ChatGPT had over 500 million registered users as of February 2025~\cite{OpenAI2025Funding}. If this trend holds, the platform could reach 1 billion users within the next 2--4 years, and 2 billion within 4--6 years, as generative AI becomes embedded in everyday devices and services.

From our experiment (see Section~\ref{Results}), we measured an average of 7.5~kB of traffic per prompt-response interaction. With the shift toward multimodal AI (text + image + video), traffic volumes may scale significantly, adding a new category of heavy Internet consumers alongside video streaming and social media platforms. %, namely images of~1-3 MB and videos of 50-200~M.

Additionally, while many users may interact infrequently, we estimate a global average of approximately 1 prompt per user per day in 2025, accounting for both active and rare users. This average is expected to increase with a Cummulative Annual Growth Rate (CAGR) of approximately 500\%, which translates into 100 and 1,000 queries per day in medium and long-term forecasts, respectively, as agents are adopted in productivity tools, operating systems, and personal devices.

Based on these growth trends, Tabl.~\ref{tab:ai_traffic_estimates} shows the following impact on traffic over the next several years.
\begin{table}[!htbp]
    \centering
    \caption{Traffic estimates in the short, medium, and long term for using AI agents}
    \begin{tabular}{|cccc|}
    \hline
         & Short Term & Medium Term & Long Term \\
         & [1-2 yr] & [2-4 yr] & [4-6 yr] \\

    \hline
    %Number of Agents &  &  &  \\
    Number of Users & 500M & 1B & 2B  \\ 
    Avg. Response [Bytes] & 7.5~kB & 1~MB & 50~MB   \\
    Usage [queries/user/day] & 1 & 100 & 1,000 \\
    \hline
    Total Monthly traffic & 3.75~TB & 100~PB & 100~EB \\
    \hline
    \end{tabular}
    \label{tab:ai_traffic_estimates}
\end{table}

The total traffic exchanged on the Internet is expected to reach around 400~EB per month for 2025 (i.e. approximately 5 Zettabytes per year). Thus, as shown in Table~\ref{tab:ai_traffic_estimates}, the IoA introduce hundreds of Exabytes of traffic into the Internet in the next years, comprising an important portion over the total. Network operators and telecommunications infrastructure need to get prepare for this AI traffic boom.

\section{Summary and discussion}
\label{sec:conclusion}

This article aims to explore the potential impact of AI agents on network and Internet traffic volumes. 

The adoption of Generative AI tools, including text-based assistants, image generation, and video creation, has demonstrated unprecedented growth, surpassing even the rapid adoption rates of the last decade's technological innovations such as social networks, the Internet of Things, and video-on-demand services.

We project that interactions between cloud-based AI agents and users will contribute significantly to overall network traffic. For text-based agents, this increase is expected to be comparable to current email and web browsing volumes. 

However, for multimodal agents capable of processing and generating multimedia content, the impact could be substantially higher, potentially adding 10 to 25\% to existing traffic levels. Moreover, in scenarios where AI agents interact with other AI agents, the traffic increase could be even more dramatic.

This surge in data exchange poses both challenges and opportunities for network infrastructure and management, necessitating proactive planning and adaptation strategies.

\section*{Acknowledgment}
The authors would like to acknowledge the support of Spanish projects ITACA (PDC2022-133888-I00) and 6G-INTEGRATION-3 (TSI-063000-2021-127) and EU project ALLEGRO (grant no. 101092766).

\bibliographystyle{IEEEtran}
\bibliography{references}
\end{document}